\documentclass[12pt]{article}

\usepackage{geometry}
\RequirePackage{amsthm,amsmath,amsfonts,amssymb,bm, xcolor,graphicx,mathrsfs}
\RequirePackage{multirow,epstopdf}
\usepackage{afterpage}
\usepackage{natbib}
\usepackage{verbatim}
\usepackage{longtable}
\usepackage{threeparttable}
\usepackage{color}
\usepackage{amsmath}
\usepackage{booktabs}
\usepackage{setspace}
\usepackage{lipsum}

\numberwithin{equation}{section}

\renewcommand{\hat}{\widehat}

\makeatother

\newcommand{\bX}{\mathbf{X}}

\newcommand{\bx}{\mathbf{x}}

\newcommand{\bbeta}{\bm{\beta}}

\newdimen\biblioindent    \biblioindent=30pt

\def\beq{\begin{equation}}
\def\eeq{\end{equation}}
\def\beqn{\begin{eqnarray}}
\def\eeqn{\end{eqnarray}}
\def\beqnn{\begin{eqnarray*}}
\def\eeqnn{\end{eqnarray*}}

\usepackage{color}

\title{A Short Note of Comparison between Convex and Non-convex Penalized Likelihood}
\begin{document}
\doublespacing
\author{Kasy Du}
\date{}
\maketitle
\begin{abstract}
This paper compares convex and non-convex penalized likelihood methods in high-dimensional statistical modeling, focusing on their strengths and limitations. Convex penalties, such as LASSO, offer computational efficiency and strong theoretical guarantees, but often introduce bias in parameter estimation. Non-convex penalties, such as SCAD and MCP, reduce bias and achieve oracle properties but pose optimization challenges due to non-convexity. The paper highlights key differences in bias-variance trade-offs, computational complexity, and robustness, offering practical guidance for method selection. It concludes that the choice depends on the problem context, balancing accuracy and computational feasibility.
\end{abstract}
\section{Background}
Let $\{(\mathbf{x}_i^T, Y_i),i=1,2,...,n\}$ be a sample from density $f\left(\mathbf{x}_{i}, Y_{i}, \bm{\beta}_0\right)$ with $\mathbf{x}_i\in\mathbb{R}^p$ and $Y_i\in\mathbb{R}$.  $\bm{\beta}_0=(\beta_{10},\beta_{20},\cdots, \beta_{p0})^\top\in\mathbb{R}^p$ is unknown parameter. Denote $l_i(\bm{\beta})=\log f\left(\mathbf{x}_{i}, Y_{i} , \boldsymbol{\beta}\right)$ and $L(\bm{\beta})=-\sum_{i=1}^{n}l_i(\bm{\beta})$. $L(\bm{\beta})$ is said to be the negative log likelihood function of the data, which could be an appropriate loss function for estimating $\bbeta_0$. We always estimate the parameter $\bbeta_0$ by minimizing the negative log likelihood function. As an example, when $f$ is a normal distribution function, the negative log likelihood function is commonly known as least square loss such that $L(\bbeta)=||Y-\bX\bbeta||^2$, where $Y$ is a $n\times 1$ column vector stacked by $Y_i$ and $\bX$ is a $n\times p$ matrix stacked by $\bx_i^T$. The model with normal distribution assumption is known as linear model.

When under low dimensional setting, the negative log likelihood function is usually convex and could be directly optimized. However, sparsity assumption is always made on the regression coefficients that assuming only a small number of $\beta_{0j}$'s are not zero. Model selection procedure is required when the parameter is sparse. In high dimensional setting, sparsity assumption is even necessary because high dimensional model parameters are usually not identifiable without sparse assumption.  To explain that, consider a linear model with $p$ much greater than $n$, then $\bX$ is no longer of full column rank. Under this case, minimizing $L(\bbeta)=||Y-\bX\bbeta||^2$ will leads to infinite number of solutions.

Under sparsity assumption, we not only want a consistent estimator of $\bbeta_0$ but also want it to be sparse. A majority part of model selection methods are based on regularization.  Define penalized likelihood as following:
\begin{eqnarray}\label{eqn231}
Q(\bm{\beta},\lambda)=\frac{1}{2n}L(\bm{\beta}) + \sum_{j=1}^{p} p_{\lambda}\left(\left|\beta_{j}\right|\right),
\end{eqnarray}
where $p_\lambda(\cdot)$ is a penalty function. $p_\lambda(\cdot)$ could take several forms. The most famous one might be the Lasso \citep{tibshirani1996regression} penalty function, where $p_\lambda(t)=\lambda|t|$.

Regularization is a fundamental technique in statistical modeling and machine learning, playing a crucial role in both variable selection and statistical inference, see \citet{tibshirani1996regression} and \citet{hoerl1970ridge}. By imposing penalties on model complexity, it helps to enhance model generalization, prevent overfitting, and improve interpretability, see \citet{hastie2005elements}. In the context of machine learning, regularization techniques are widely used in model fitting to ensure that the model remains robust and performs well on unseen data \citep{ng2004feature, tian2022comprehensive, ma2024machine, chen2019comparison, cui2024hypothesis, ci2023proactive, ci2023gfpose, cui2023enhancing, kotsilieris2022regularization, ci2019optimizing}. In addition, regularization has significant applications in biostatistics, where it is applied to problems such as genomic data analysis, survival modeling, and clinical trials, where the number of predictors often exceeds the sample size \citep{friedrich2023regularization, hastie2004efficient, vieira2023comparison, hafemeister2019normalization, wang2023variable,wang2022normalized, zhang2022stochastic}.  Whether through Lasso for variable selection or Ridge for controlling the magnitude of coefficients, regularization is essential in refining models across various applications \citep{zou2005regularization,tibshirani1996regression}.

Since $p_\lambda(t)=\lambda|t|$ is a convex function, when we apply the Lasso penalty, optimizing (\ref{eqn231}) is still a convex problem. However, some authors have pointed out disadvantages of the Lasso and tried several other penalty functions, see the SCAD \citep{fan2001variable}, the MCP \citep{zhang2010nearly}, and the Dantzig selector \citep{candes2007dantzig}, among others. Nonfolded concave penalties have broad applications. \cite{liang2010estimation} and \cite{cui2024estimation} applied it to semi-parametric models. \cite{cui2024double} utilized the unfolded concave penalty to nonlinear models and models with measurement error. The SCAD penalty is given by 
$$p_{\lambda}^{S C A D}\left(\beta\right)=\left\{\begin{array}{ll}
\lambda\left|\beta\right| & \text { if }\left|\beta\right| \leq \lambda ,\\
-\left(\dfrac{\left|\beta\right|^{2}-2 a \lambda\left|\beta\right|+\lambda^{2}}{2(a-1)}\right) & \text { if } \lambda<\left|\beta\right| \leq a \lambda, \\
\dfrac{(a+1) \lambda^{2}}{2} & \text { if }\left|\beta\right|>a \lambda,
\end{array}\right.$$
with some specific $a>2$. 

The MCP penalty is given by
$$p(\beta)=\left\{\begin{array}{ll}
\lambda|\beta|-\dfrac{\beta^{2}}{2 a} & \text { if }|\beta| \leq a \lambda, \\
\dfrac{a \lambda^{2}}{2} & \text { otherwise, }
\end{array}\right.$$ with $a>1$.  

\begin{figure}
    \centering
    \includegraphics[width=3in]{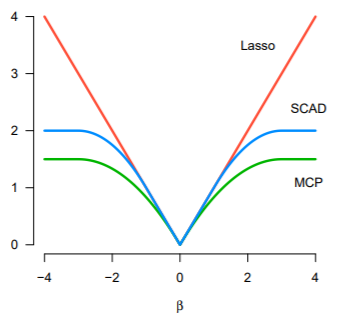}
    \caption{Some penalty functions.}
    \label{fig1}
\end{figure}
Figure \ref{fig1} depicts different penalty functions. We could see that the SCAD and the MCP penalties are non-convex, differently with the Lasso penalty. 

As we know, optimizing of non-convex function is always harder. Why do we need to introduce the Lasso penalty? Does it worth doing that? In this project I want to firstly discuss about  main disadvantages and advantages of various penalized method. Then I want to compare main methods and algorithms  of optimizing penalized likelihood under various penalized functions. At last I want to show their applications in a real world example.

\section{Properties of Different Penalties}
The Lasso owns its popularity to its convexity and computational properties in statistical learning and modeling. However several authors such as \citet{zou2005regularization} and \citet{zhao2006model} have pointed out in theory that the Lasso penalty always introduces a bias on its large components without sacrificing model selection consistency. The SCAD and the MCP penalties can largely avoid it. The idea of the SCAD and the MCP penalties come from observation of the first order derivative of the penalty functions.  
The first order derivative of the Lasso, the SCAD and the MCP penalty is given by
\begin{itemize}
    \item Lasso: $p^\prime_{\lambda}(t)=\lambda$
    \item SCAD: $p_{\lambda}^{\prime}(t)=\lambda\left\{I(t \leq \lambda)+\dfrac{(a \lambda-t)_{+}}{(a-1) \lambda} I(t>\lambda)\right\} \quad \text {, for some } a>2 $
    \item MCP: $p_{\lambda}^{\prime}(t)=(\lambda-t / a)_{+}$
\end{itemize}

\begin{figure}
    \centering
    \includegraphics[width=3in]{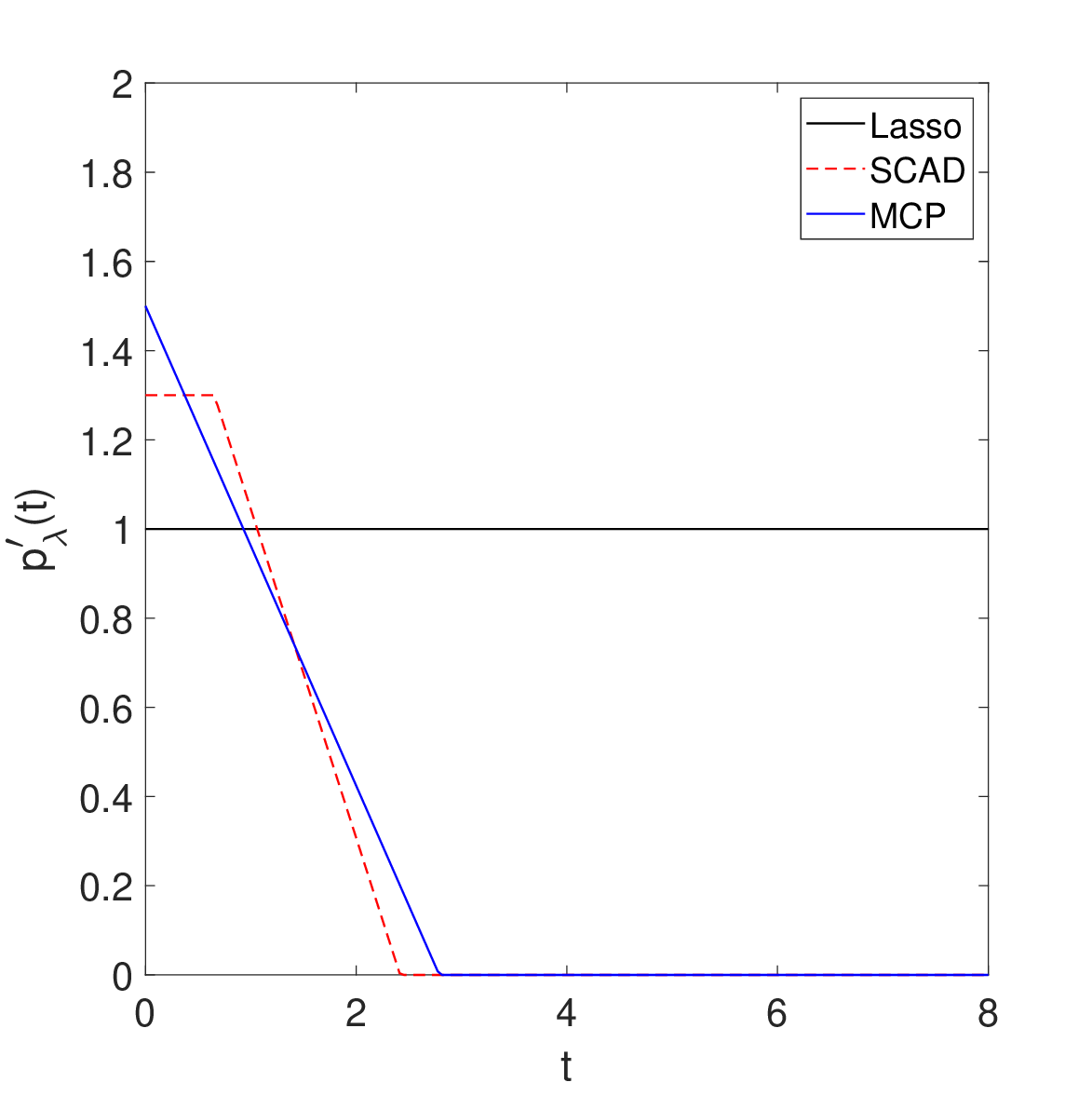}
    \caption{Derivative of some penalty functions.}
    \label{fig3}
\end{figure}

Figure \ref{fig3} depicts the derivative of these penalty functions. From the plot we can see that the Lasso penalty gives equal penalty to any value of the parameters, no matter large or small. Recalling that our motivation of adding penalties is to make the estimator to be sparse, which means that we punish on small parameters to be shrunk to  zero. However, meanwhile that will also lead to a large bias on estimating large parameters. Therefore we do not want to give equal penalty on different parameters. Ideally, we want to give penalty only on the estimator of coefficients close to zero. From Figure \ref{fig3}, we can see that the SCAD and the MCP utilize the idea such that the penalty is positive only when $t$ is small and becomes zero when $t$ is large.

To illustrate how the non-concave penalty works, we design a toy experimental example to show it. 

\emph{Model 1:} Our model is generated from $Y=X\beta+\epsilon$, $n=200:$ $p=1000:$  $X\in\mathbb{R}^{200\times1000} $ is a fat matrix. $\beta\in\mathbb{R}^{1000\time 1}$ $\epsilon_i\sim N(0,1)$. We set $\beta_1,\beta_2,\cdots, \beta_{10}=1, \beta_j=0$, when $j>10.$ Ground truth of sparsity level is 10. We replicate for 100 times. MSE of $\hat{\bbeta}$ is computed from the average of $||\hat{\bbeta}-\bbeta||$ over 100 replications.

\begin{figure}[h]
\begin{minipage}[t]{0.5\linewidth}
\begin{center}
\includegraphics[width=2.5in, height=2in]{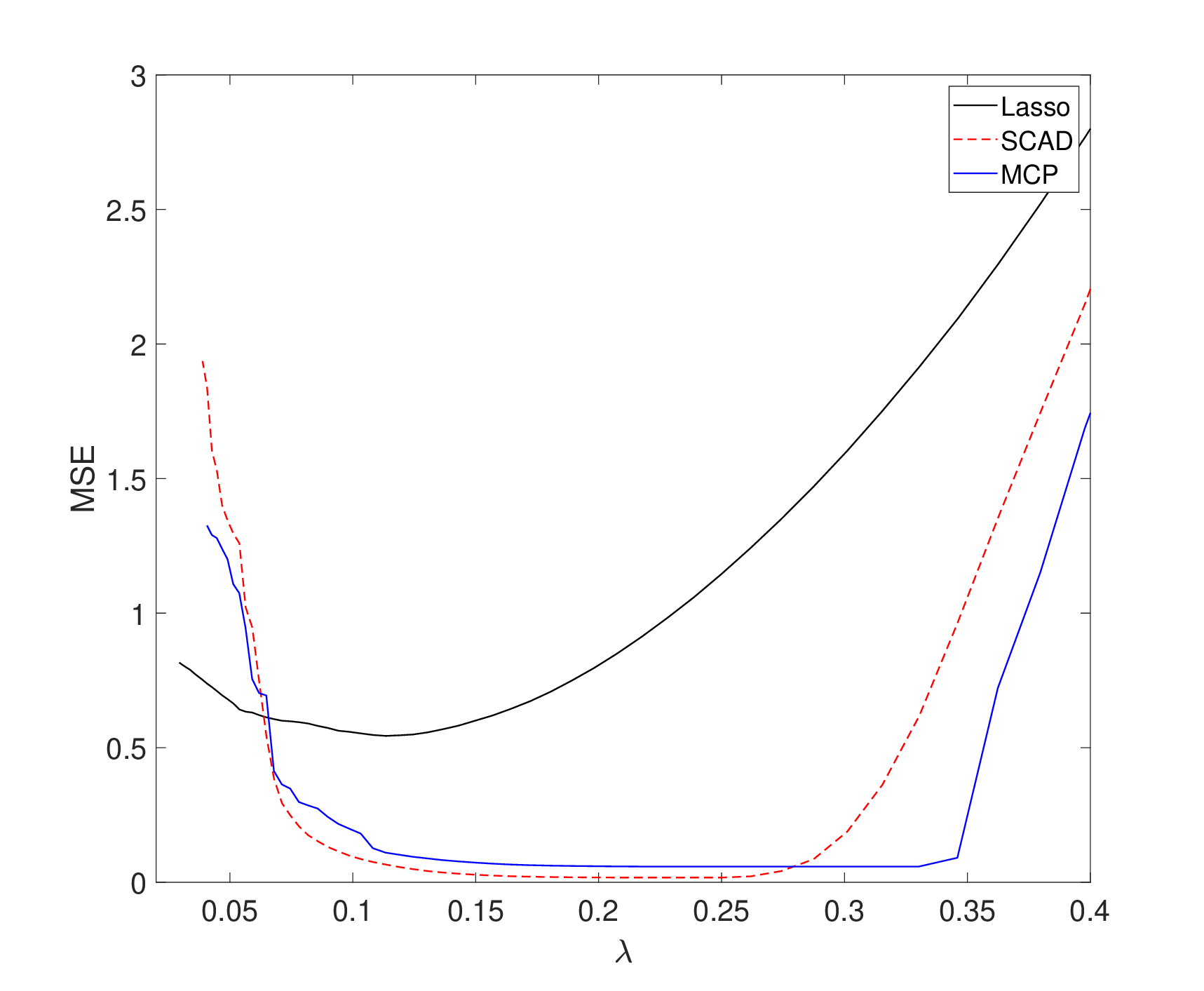}
\end{center}
\end{minipage}%
\begin{minipage}[t]{0.5\linewidth}
\begin{center}
\includegraphics[width=2.5in, height=2in]{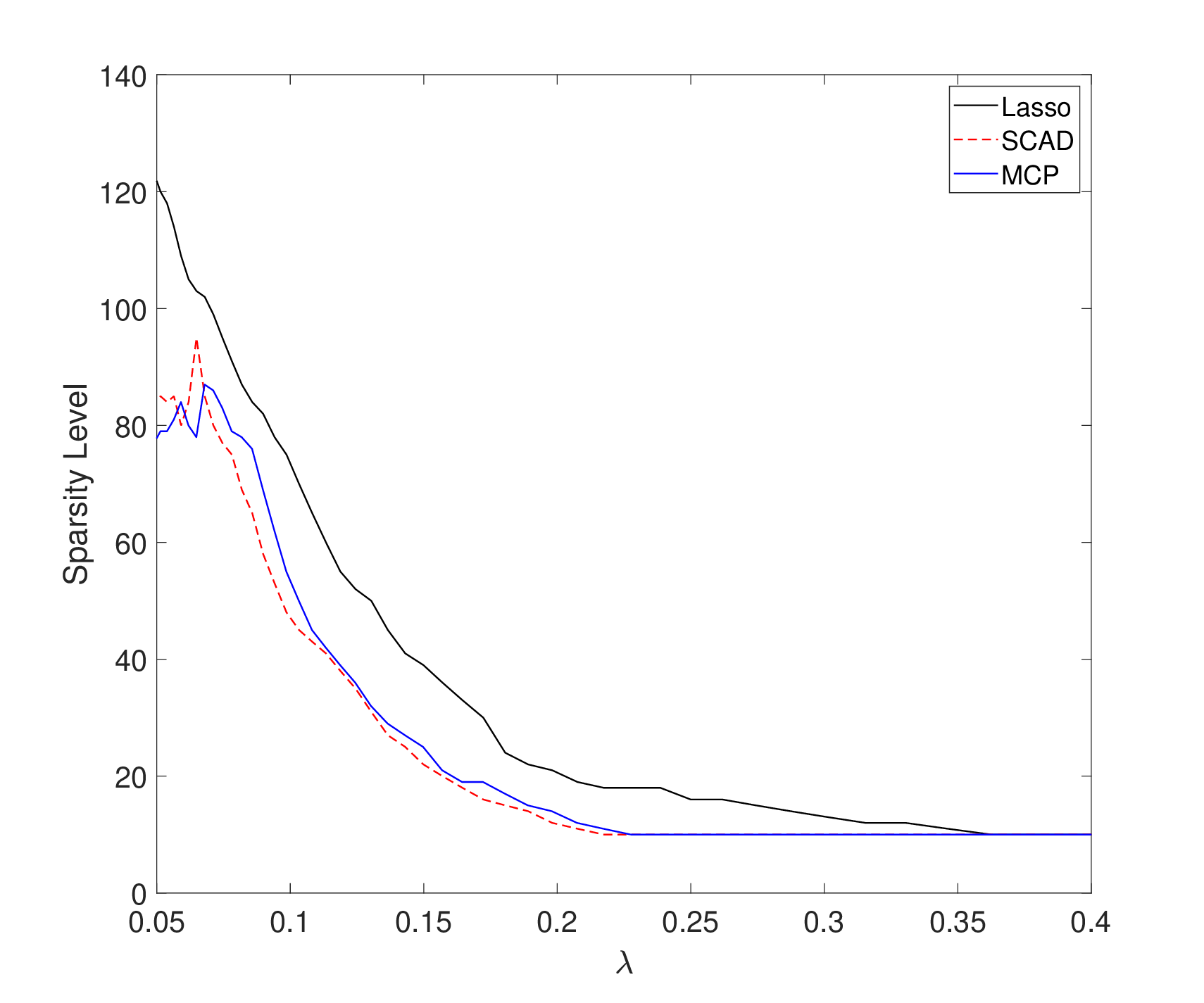}
\end{center}
\end{minipage}\caption{\emph{Left: MSE of different penalties under different $\lambda$ Right: Sparsity level of different penalties under different $\lambda$. }}\label{Fig2}
\end{figure}

Figure \ref{Fig2} shows the fitting result of the model. Firstly noticing that from the left plot of Figure \ref{Fig2}, the MSE of the Lasso estimator achieves minimum at $\lambda=0.12$. However, the sparsity level of the Lasso estimator under $\lambda=0.12$ is pretty high and is far from the ground truth 10. Sparsity level of the Lasso estimator can achieve ground truth when $\lambda$ is greater than 0.35, but the MSE is high when $\lambda$ is large. It 
indicates that the Lasso cannot approximate the sparsity level and the parameters well at the same time. On the other hand, the SCAD and the MCP penalties can give us an estimator with ground truth sparsity level without sacrificing estimation accuracy.

\section{Algorithms and Experimental Results}
For the SCAD and the MCP penalties, most of the authors do not talk much about algorithms they used for optimization, especially the performance of convergence, such as escape time and frequency of convergence. In this section I derive the formula of coordinate decent algorithm on loss function with different penalty function. For simplicity, we only consider the case under linear model.

Coordinate decent is a common algorithm we could consider to optimize both convex or non-convex problem. At each iteration, we update one coordinate while fixing the others. Fortunately, for the Lasso, the SCAD  and the MCP penalty, coordinate decent algorithm gives us closed form for each iteration $\beta^{(k)}_j=S(\bX_j^T(Y-\bX_{-j}\bbeta^{(k-1)}_{-j}),\lambda)$ with different $S$:
    \begin{itemize}
        \item    Lasso: $$S(z,\lambda)=\operatorname{sgn}(z)(|z|-\lambda).$$
        \item SCAD:
        $$ S(z,\lambda)=\left\{\begin{array}{ll}
\operatorname{sgn}(z)(|z|-\lambda)_{+}, & \text {when }|z| \leq 2 \lambda \\
\operatorname{sgn}(z)[(a-1)|z|-a \lambda] /(a-2), & \text { when } 2 \lambda<|z| \leq a \lambda \\
z, & \text { when }|z| \geq a \lambda
\end{array}\right.
        $$
        \item MCP:
        $$S(z,\lambda)=\left\{\begin{array}{ll}
\operatorname{sgn}(z)(|z|-\lambda)_{+} /(1-1 / a), & \text { when }|z|<a \lambda ; \\
z, & \text { when }|z| \geq a \lambda
\end{array}\right.$$
    \end{itemize}

\begin{figure}
    \centering
    \includegraphics[width=3in]{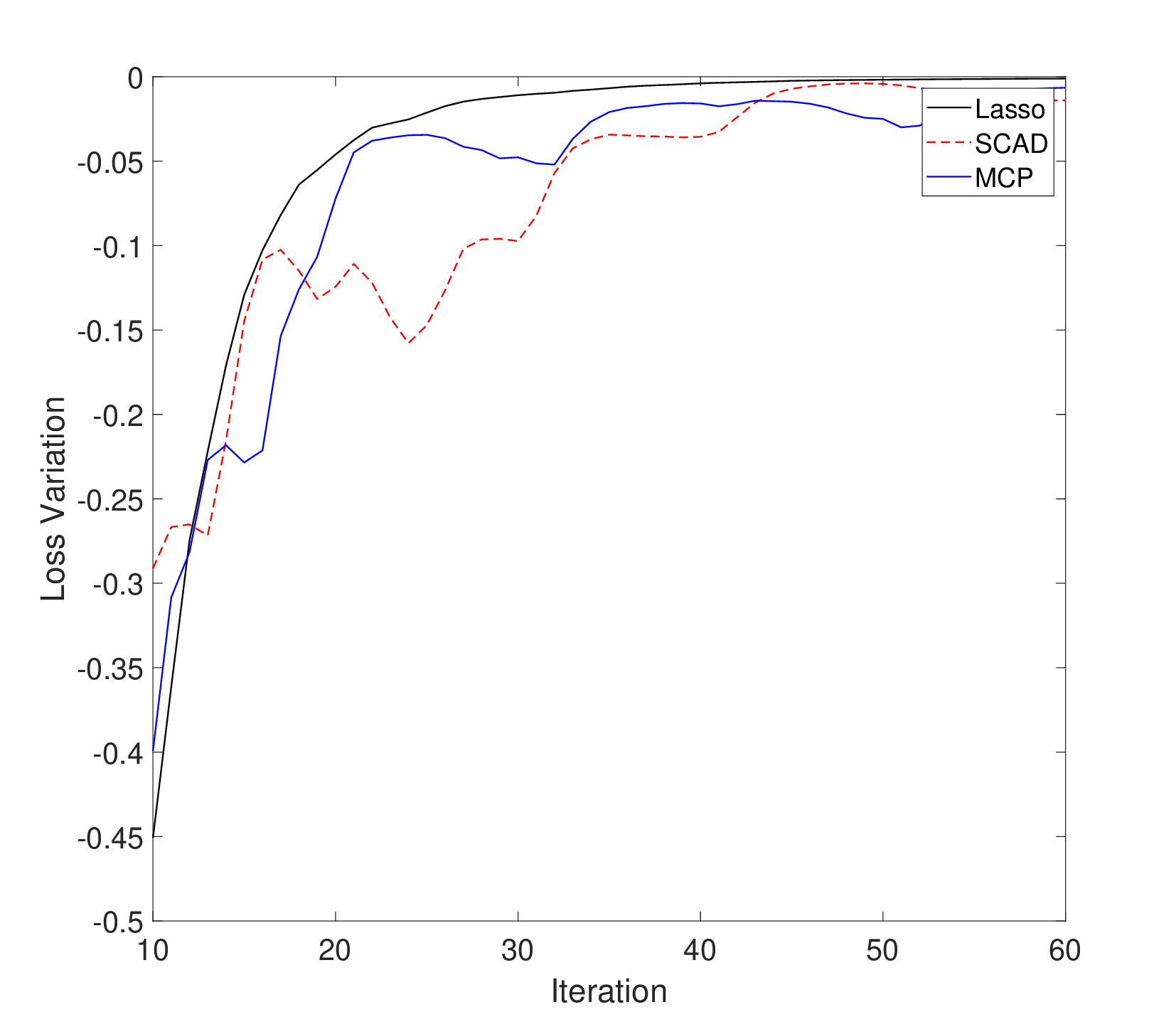}
    \caption{Loss Variation.}
    \label{fig4}
\end{figure}

We still consider Model 1 in section 2. We fit the model with coordinate decent algorithm and record the loss variation in each iteration. Figure \ref{fig4} depicts loss variation in each iteration. (The first ten iterations are removed)

We could see from Figure \ref{fig4} that compared with the Lasso, convergence of the SCAD and the MCP penalties
is not stable. What is more, Table \ref{tab1} shows the average time required for convergence of the three types of aforementioned penalized loss functions. We could see that the SCAD and MCP penalties converge much slower than the Lasso. It is not surprised to see this because optimization of non-convex functions is usually harder and more time consuming.

\begin{table}[]
\centering
\caption{Average Time of Convergence}\label{tab1}

\begin{tabular}{lll}
\toprule
Lasso & SCAD & MCP \\ \hline
31s   & 86s  & 92s \\ \toprule
\end{tabular}
\end{table}
\section{Real World Example}
We now illustrate the proposed methodology by an application to the gene
expression data set consisting of expression levels of 200 genes from 120 rats.
This data set is an extraction from a study that aims to find gene regulation
and variation for relevant human eye disease. Expression quantitative trait
locus (eQTL) mapping in laboratory rats are used because of the similarity of
rat and human genes, see \cite{scheetz2006regulation} for details. The response
variable is the expression level of the TRIM32 gene, which is linked to the
BardetBiedl syndrome. In this example, we apply loss function with the Lasso and the SCAD penalties to the data set and aim to locate the genes, which are highly
correlated with gene TRIM32.
We standardize the data to make each variable
with zero mean and unit standard deviation.

\begin{figure}[h]
\begin{minipage}[t]{0.5\linewidth}
\begin{center}
\includegraphics[width=2.5in, height=2in]{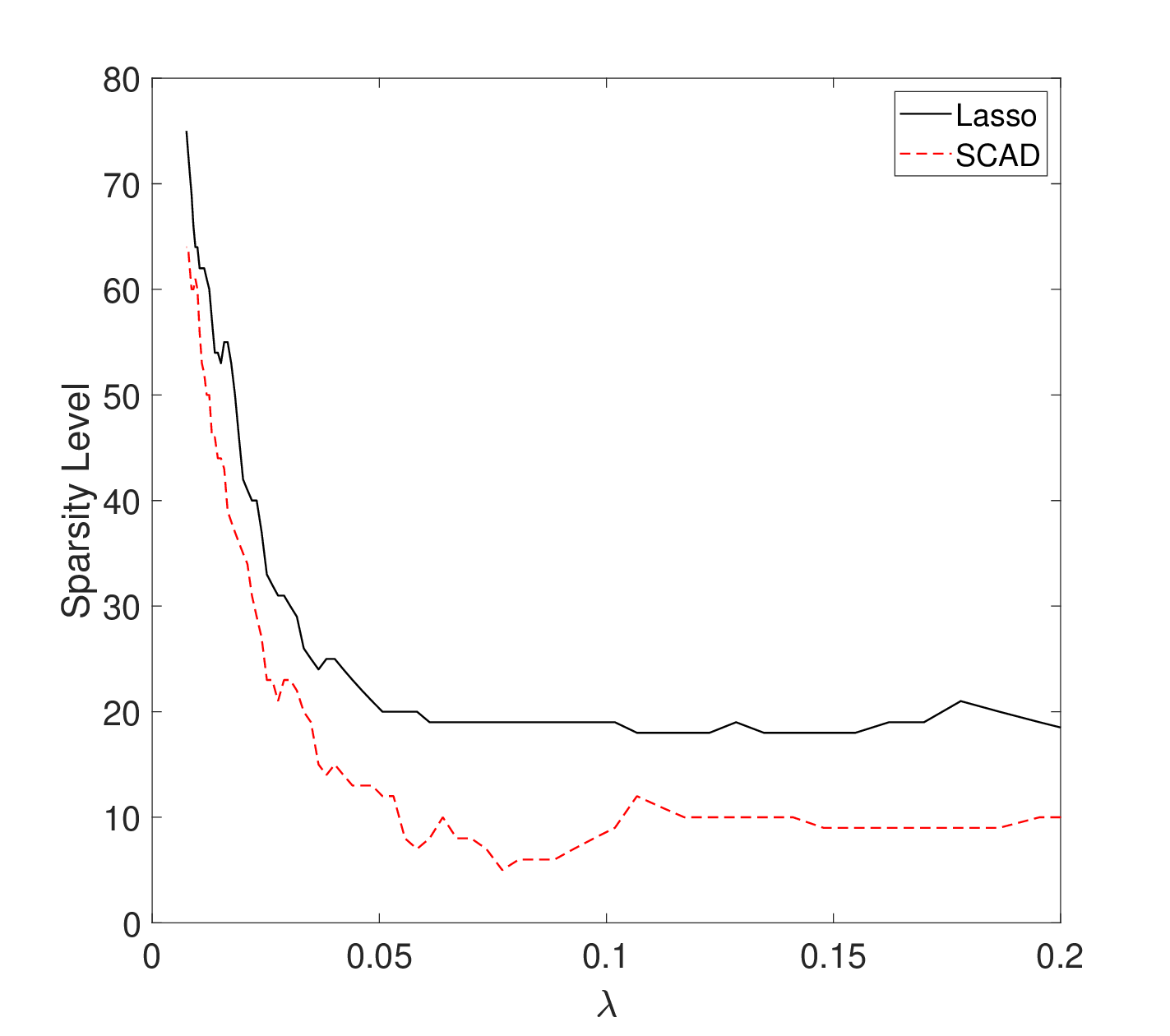}
\end{center}
\end{minipage}%
\begin{minipage}[t]{0.5\linewidth}
\begin{center}
\includegraphics[width=2.5in, height=2in]{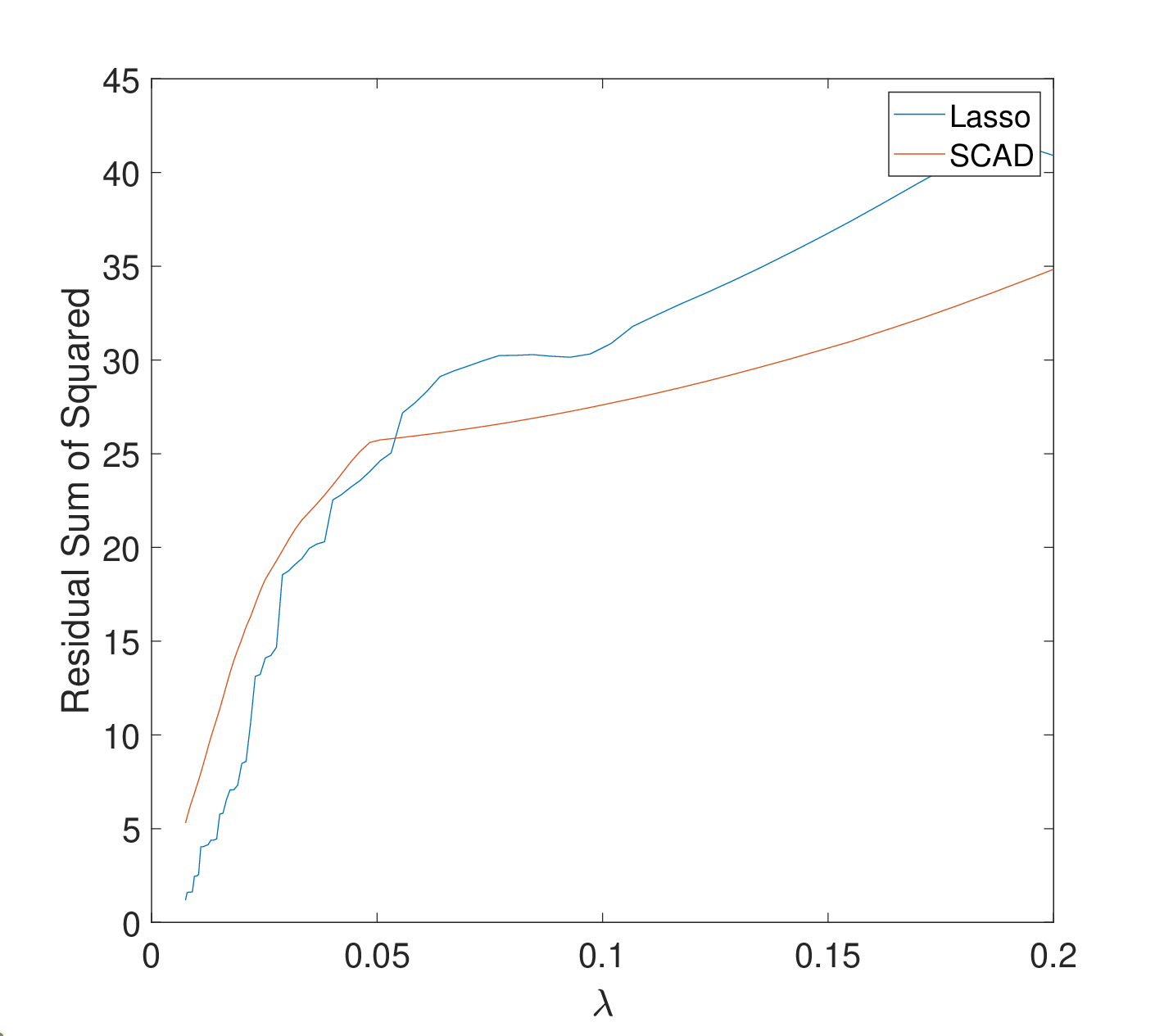}
\end{center}
\end{minipage}\caption{\emph{Left: Sparsity level Comparison. Right: Residual sum of squares comparison. }}\label{Fig5}
\end{figure}

Figure \ref{Fig5} shows the model fitting results. We could see from the left panel of Figure \ref{Fig5} that the SCAD penalty tends to provide a more sparse solution. 
The left panel shows the residual sum of squares to illustrate how well the model fitting is. We apply 10 fold cross validation to select $\lambda$. For the Lasso, it gives $\lambda=0.146,$ and for the SCAD, it gives $\lambda=0.158$ We could see from the right penal that under the selected $\lambda$, the SCAD performs better than the Lasso penalty.

\bibliographystyle{apalike}
\bibliography{bibliography}

\end{document}